\documentclass[]{iptconf}

\microtypesetup{expansion=false}

\usepackage[most]{tcolorbox}
\usepackage{tabularray}
\UseTblrLibrary{booktabs}
\usepackage{dsfont}
\usepackage{mathrsfs}
\usepackage{wrapfig}
\usepackage[version=4]{mhchem}
\let\babelch\ch
\usepackage{chemformula}

\let\ch\babelch
\usepackage{forest}
\usepackage{tikz}
\usepackage{pgfplots}
\pgfplotsset{compat=1.18}
\usepackage{listings}
\usetikzlibrary{shadows,arrows.meta}

\newcommand{\be}{\begin{equation}}
\newcommand{\ee}{\end{equation}}
\newcommand{\bea}{\begin{eqnarray}}
\newcommand{\eea}{\end{eqnarray}}

\usepackage[backend=biber]{biblatex}
\title{TRACE ANOMALY, EFFECTIVE DEGREES OF FREEDOM, AND CHEMICAL-POTENTIAL EFFECTS NEAR THE QCD CROSSOVER}

\author[y.kryvenko-emetov@kpi.ua]{Y.~D.~Krivenko-Emetov}{1,2,3}

\affiliation{National Technical University of Ukraine, 03056, Kyiv}{1}
\affiliation{Institute for Nuclear Research of the NAS of Ukraine, 03680, Kyiv}{2}
\affiliation{Taras Shevchenko National University of Kyiv, Kyiv, Ukraine}{3}
\vspace{0.5cm}


\abstract{%
A compact analytical scheme is presented for describing ultra-dense hadronic matter,
which combines a multicomponent van der Waals (vdW)-type description
with temperature-dependent effective degrees of freedom.
Although the vdW formalism successfully reproduces interactions at finite density,
in its standard form it cannot describe lattice-QCD thermodynamics,
since it uses a fixed degeneracy.

It is shown that a consistent description of the equation of state requires
a temperature-dependent degeneracy $g(T)$ and an effective chemical
potential $\mu(T)$. Within this approach, the trace anomaly (the trace of the energy-momentum tensor),
i.e. the measure of nonconformality of the energy-momentum tensor normalized to $T^4$,
is naturally reproduced together with its peak structure near the crossover region.

The effective chemical-potential sector becomes particularly important
in baryon-rich matter, whereas for the mesonic sector
a separate dynamical description of the degrees of freedom is required.
}

\begin{document}
\PaperLanguage{english}
\maketitle
\selectlanguage{english}

\pgfplotsset{compat=1.18}

\section{Introduction}

Relativistic nucleus--nucleus collisions create a short-lived strongly interacting system that
evolves from a deconfined strongly interacting quark-gluon plasma~\cite{Shuryak2004, Gyulassy2005} to a hadronic fireball~\cite{Shuryak2004, BraunMunzinger2007}.
At intermediate stages such a system can be regarded as a set of quasi-equilibrium subsystems,
which admits a thermodynamic description~\cite{Krivenko2023}.

The van der Waals (vdW) approach gives a simple analytical scheme that takes into account
short-range repulsion~\cite{Gorenstein1999, Rischke1991} and attraction in a multicomponent gas at intermediate distances~\cite{Vovchenko2017, Krivenko2017Attraction, Krivenko2019}.
However, in its standard form it assumes a fixed number of degrees
of freedom and therefore cannot reproduce lattice-QCD thermodynamics~\cite{Borsanyi2014, Bazavov2014}.

In the quark-gluon plasma regime an important role is played by
the effects of screening of color interactions (Debye screening)
and by the quasiparticle structure of the medium, which leads
to temperature-dependent effective masses and degrees of freedom.

The approach proposed in this work should be understood as an effective phenomenological model.
It is not derived directly from first-principles QCD, but is constructed so as
to capture the dominant thermodynamic effects associated with
temperature-dependent activation of degrees of freedom and interaction effects
in strongly interacting matter (for example, in the multicomponent model of a nuclear fireball at the freeze-out stage~\cite{Krivenko2023}).

The central hypothesis of the present work is that the QCD trace anomaly
can be interpreted as a direct consequence of the interplay between
temperature-dependent effective degrees of freedom and the effects
of quantum degeneracy.

We show that:
\begin{itemize}
\item the Hagedorn growth of the hadronic spectrum~\cite{Hagedorn1965},
\item the emergence of quantum degeneracy,
\item and interaction-induced shifts of the chemical potential,
\end{itemize}
together provide a unified and analytically transparent description of the QCD crossover,
consistent with lattice data.
A preliminary version of this work was presented
at the XXVI Annual Scientific Conference of the Institute
for Nuclear Research of the NAS of Ukraine.

\section{Basic scheme: interacting hadronic matter at high density}

We use an effective thermodynamic description built on two
main ingredients:
\begin{itemize}
\item van der Waals (vdW)-type interactions,
\item temperature-dependent effective degrees of freedom.
\end{itemize}

At low densities the van der Waals equation reduces to
\begin{equation}
P(T,n) \simeq T n \left[1+B(T)n\right],
\end{equation}
which corresponds to the second virial correction and is equivalent to an ideal gas
with a shifted chemical potential
\begin{equation}
\mu_{\mathrm{int}} = \mu - T B(T)n.
\label{mu_int}
\end{equation}

For baryonic matter a minimal two-component structure
(protons and neutrons) is introduced:
\begin{equation}
P =
\sum_i \frac{T n_i}{1-b_{ii}n_i-\tilde b_{ij}n_j}
-
\sum_{i,j} a_{ij}n_i n_j.
\end{equation}

Although this scheme takes interaction effects into account, it does not describe
lattice-QCD thermodynamics. This indicates that the number of effective degrees of freedom
must depend on temperature.

A detailed account of the attractive component of the interaction in a multicomponent
hadron gas in the grand canonical ensemble was considered in
\cite{Krivenko2017Attraction},\cite{Krivenko2019}.

\subsection{Degeneracy temperature of the hadronic component with interaction effects}

Near the Hagedorn temperature the hadronic medium can no longer be described
as a classical gas. The corresponding criterion of quantum degeneracy is
\begin{equation}
n_h\lambda_T^3\sim 1,
\end{equation}
where
\begin{equation}
\lambda_T=\sqrt{\frac{2\pi\hbar^2}{m_hT}}
\end{equation}
is the thermal de Broglie wavelength.

This determines the degeneracy temperature of the hadronic component:
\begin{equation}
T_{\mathrm{deg}}^{(h)}
\simeq
\frac{2\pi\hbar^2}{m_h}
\left(\frac{n_h}{\zeta(3/2)}\right)^{2/3}.
\end{equation}

In the presence of interactions the density changes due to excluded-volume
or mean-field effects. In the simplest approximation
\begin{equation}
n_h(T)=n_h^{(0)}(T)\exp[-B(T)n_h],
\end{equation}
where $B(T)$ is an effective interaction parameter.

Substituting this expression into the degeneracy criterion, we obtain
\begin{equation}
T_{\mathrm{deg}}^{(h,\mathrm{int})}
\simeq
T_{\mathrm{deg}}^{(h)}
\exp\left[-\frac{2}{3}B(T)n_h\right].
\end{equation}

Thus, repulsive interactions ($B>0$) suppress effective
degeneracy, whereas attractive interactions ($B<0$) enhance it.

An additional effect arises due to the Hagedorn growth of the hadronic
spectrum,
\begin{equation}
\rho(m)\sim \exp(m/T_H),
\end{equation}
which leads to a rapid growth of $n_h(T)$ as $T\to T_H$.
As a result, the hadronic component approaches the quantum-degenerate
regime already in the crossover region.

\subsection{Degeneracy temperature of the quark-gluon plasma with interaction effects}

For an ultrarelativistic quark gas at finite baryon
chemical potential, the quark chemical potential is
\begin{equation}
\mu_q=\frac{\mu_B}{3}.
\end{equation}

In the approximation of a degenerate Fermi gas, the characteristic Fermi scale determined by the density
of the ultrarelativistic Fermi gas can be estimated as

\begin{equation}
T_{\mathrm{deg}}^{(0)}=
\left[\mu_q^3+\pi^2\mu_qT^2\right]^{1/3}.
\end{equation}

Taking interactions into account, the chemical potential is split into kinetic
and mean-field parts:
\begin{equation}
\mu_q=\mu_q^*+U_v(n,T),
\end{equation}
where $U_v$ is the vector mean-field potential.

Then
\begin{equation}
\mu_q^*=\mu_q-U_v(n,T),
\end{equation}
and the degeneracy temperature takes the form
\begin{equation}
T_{\mathrm{deg}}^{\mathrm{int}}=
\left[(\mu_q^*)^3+\pi^2\mu_q^*T^2\right]^{1/3}.
\end{equation}

For a repulsive interaction
\begin{equation}
U_v>0
\end{equation}
we have
\begin{equation}
T_{\mathrm{deg}}^{\mathrm{int}}<T_{\mathrm{deg}}^{(0)}.
\end{equation}

This reflects the fact that part of the chemical potential is spent
on the interaction energy, rather than on filling the Fermi levels.

For the baryon-rich region of the phase diagram in the crossover region, for example
\[
T\sim 160~\mathrm{MeV},\qquad \mu_B\sim 900~\mathrm{MeV},
\]
we obtain
\begin{equation}
T_{\mathrm{deg}}^{\mathrm{int}}\sim400~\mathrm{MeV},
\end{equation}
that is
\begin{equation}
T_{\mathrm{deg}}^{\mathrm{int}}>T.
\end{equation}

The quantity \(T_{\rm deg}\) should be understood as an effective
characteristic Fermi scale rather than a strict phase-transition temperature.

Thus, for such parameters the quark component is in a regime
of substantial Fermi degeneracy even
near the QCD crossover.

\subsection{Limitation of fixed degeneracy}

The key limitation of the standard vdW approach is the assumption of
a fixed number of degrees of freedom. In QCD matter the effective number
of degrees of freedom changes with temperature:
\begin{equation}
\varepsilon \sim g(T)T^4.
\end{equation}

Therefore a consistent description requires a temperature-dependent degeneracy.

\section{Trace anomaly (trace of the energy-momentum tensor) and the emergence of effective degrees of freedom}

The key observable quantity is the trace anomaly~\cite{Bazavov2014, Borsanyi2014}.

In QCD the trace anomaly has a microscopic origin
associated with the breaking of scale invariance:

\[
T^\mu_{\ \mu} = \frac{\beta(g)}{2g} F^a_{\mu\nu} F^{a\mu\nu}.
\]

Therefore the quantity
\[
\Delta(T) = \frac{\varepsilon - 3P}{T^4}
\]
is directly related to the gluon-field condensate.

Within the proposed phenomenological model this effect
is effectively reproduced through the temperature dependence of the
degrees of freedom \( g(T) \) and the effective chemical potential \( \mu(T) \).

If the nonconformal part of the equation of state changes rapidly with temperature~\cite{Kapusta2006},
then this is directly reflected in the quantity $\Delta(T)$.
Using the thermodynamic identity
\[
\varepsilon=T\frac{dP}{dT}-P,
\]
we have
\[
\Delta(T)=T\frac{d}{dT}\left(\frac{P}{T^4}\right).
\]
Thus, the maximum of \(\Delta(T)\) appears in the region where the dimensionless pressure
\(P/T^4\) changes most rapidly. In the phenomenological model this rapid change
is effectively parameterized by the temperature function \(g(T)\).
Thus, the rapid change of $g(T)$ naturally generates the peak of the trace anomaly.
This quantity directly probes interaction effects and the change of the effective
number of degrees of freedom.

\subsection{Temperature-dependent degeneracy}

We introduce an effective degeneracy $g(T)$, which interpolates between the hadronic
regime and the quark-gluon plasma regime:

\begin{equation}
\begin{aligned}
g(T)=
\begin{cases}
g_0\left[1+\alpha\left(\frac{T}{T_H}\right)^2\right], & T<T_H, \\
g_0\left[1+\alpha\left(\frac{T}{T_H}\right)^2\right] \\
\qquad + g_\infty\left[1-e^{-\beta(T-T_H)}\right], & T\ge T_H.
\end{cases}
\end{aligned}
\end{equation}

This form describes the growth of the contribution of hadronic resonances and the gradual
activation of partonic degrees of freedom. It follows from this that
the QCD crossover takes place between two regimes, both of which have
a quantum nature: an almost degenerate hadronic resonance gas and
a partially degenerate quark-gluon medium.

\subsection{Effective chemical potential}

The particular functional form was chosen as the
simplest smooth parameterization capable of reproducing
the monotonic suppression of the effective chemical
potential with temperature.

We parameterize the temperature dependence of the
chemical potential as
\begin{equation}
\mu(T)= e_f-\frac{T^A}{m_{\mathrm{eff}}}.
\label{muT}
\end{equation}

Here \(e_f\) and \(m_{\mathrm{eff}}\) are phenomenological
parameters associated with the effective chemical potential
scale and the effective mass scale in the crossover region.

The parameter \(e_f\) characterizes the effective baryonic
energy scale at low temperatures, whereas
\(m_{\mathrm{eff}}\) controls the temperature dependence
of the suppression of the effective chemical potential
and therefore determines the rate of transition from
hadronic to partonic thermodynamic regimes.

The particular functional form was chosen as the
simplest smooth parameterization capable of reproducing
the monotonic suppression of the effective chemical
potential with temperature.

This parameterization should be regarded as an effective
generalization of the shifted chemical potential
appearing in the van der Waals description,
Eq.~(\ref{mu_int}),
which was previously obtained for interacting hadronic matter
within the excluded-volume and multicomponent van der Waals
approach~\cite{Gorenstein1999,Krivenko2023}.

In the low-density limit the interaction contribution
produces an effective reduction of the chemical potential,
\[
\mu_{\mathrm{int}}=\mu-TB(T)n,
\]
which reflects the suppression of available thermodynamic
states due to interaction effects.

The phenomenological form (\ref{muT}) extends this idea
to the crossover region by incorporating the gradual melting
of hadronic bound states and the transition toward partonic
degrees of freedom.
It therefore provides an effective interpolation between
the hadronic and quark--gluon regimes.

The decrease of \(\mu(T)\) with temperature imitates
the weakening of baryonic dominance due to the increasing
importance of quark--antiquark excitations and the gradual
reduction of the role of bound hadronic configurations.

The linear suppression form was chosen as the minimal smooth parameterization capable of reproducing the gradual reduction of baryonic dominance near the crossover.

Within this interpretation, the parameterization effectively describes:
\begin{itemize}
\item the hadronic regime at low temperatures,
\item the melting of bound states,
\item the gradual activation of partonic degrees of freedom,
\item the approach to an almost symmetric quark--gluon plasma (QGP).
\end{itemize}

More microscopic realizations may involve quasiparticle self-energies or Polyakov-loop induced suppression factors.

\subsection{Model of the trace anomaly}

The trace anomaly (interaction measure)
is defined in the standard way:
\[
\Delta(T,\mu)=\frac{\varepsilon-3P}{T^4},
\]
which characterizes the deviation of the system from conformal behavior.

Motivated by the general structure of the expansion
of thermodynamic quantities in even powers of the chemical potential,
we introduce the following compact phenomenological representation (in small-\(\mu/T\) expansion):

\[
\Delta(T,\mu)
=
g(T)\left[
1+B\left(\frac{\mu(T)}{T}\right)^2
\right],
\]

An alternative phenomenological parameterization
may be constructed using the mixed dimensionless
scaling variable
\[
\eta(T)=\frac{T}{T_H^2}\mu(T),
\]
which combines the thermal scale and the effective
chemical-potential sector.

In this case the trace anomaly is parameterized as
\[
\Delta(T,\mu)=
g(T)\left[
1+D\eta^2(T)
\right],
\]
or explicitly
\[
\Delta(T,\mu)=
g(T)\left[
1+D\left(\frac{T}{T_H^2}\right)^2\mu^2(T)
\right].
\]

where \(g(T)\) should be understood as an effective nonconformal spectral weight,
which accumulates the temperature activation of thermodynamically active degrees
of freedom and their contribution to the quantity \(\varepsilon-3P\), while the factor multiplying \(\mu(T)\)
takes into account the additional nonconformality associated with the baryonic
occupation of the system.

Unlike the standard small-\(\mu/T\) expansion,
this parameterization should be regarded as an
effective phenomenological scaling ansatz rather
than as a direct thermodynamic expansion.

Such a form effectively accumulates the interplay
between thermal activation and chemical-potential
effects near the crossover region.

Such a representation is equivalent to the assumption that the nonconformal part
of the equation of state can be described by an effective bag-like or mean-field
contribution:

\begin{equation}
\begin{aligned}
P(T,\mu) &= P_{\rm conf}(T,\mu)-U(T,\mu),\\
\varepsilon(T,\mu) &= 3P_{\rm conf}(T,\mu)+U(T,\mu).
\end{aligned}
\end{equation}

In this case one immediately obtains
\[
\varepsilon-3P=4U(T,\mu),
\]
which leads precisely to the form for \(\Delta(T,\mu)\) given above.

The quadratic dependence on \(\mu\) has a general character.
Since thermodynamic quantities are even functions of the
chemical potential (invariance with respect to
\(\mu\to -\mu\)), the expansion at small \(\mu/T\) has the form
\[
\Delta(T,\mu)=\Delta(T,0)
\left[
1+C_2(T)\left(\frac{\mu}{T}\right)^2
+\mathcal{O}\!\left(\frac{\mu^4}{T^4}\right)
\right].
\]
In the simplest phenomenological realization we set
\[
C_2(T)=B=\text{const},
\]
which leads to the compact parameterization
\[
\Delta(T,\mu)=g(T)
\left[
1+B\left(\frac{\mu(T)}{T}\right)^2
\right].
\]

As a result, in two regions of phase space \((\mu,T)\),
the trace anomaly can be approximately represented as an expansion
in the corresponding small parameters.

For the regime of weak baryonic occupation,
\[
\frac{\mu(T)}{T}\ll1,
\]
the natural expansion parameter is
\[
\frac{\mu(T)}{T},
\]
and the trace anomaly takes the form
\begin{equation}
\Delta(T,\mu)=
g(T)\left[
1+B\left(\frac{\mu(T)}{T}\right)^2
+\mathcal O\!\left(
\frac{\mu^4(T)}{T^4}
\right)
\right].
\end{equation}

Thus, the QCD crossover may be analyzed in two complementary
asymptotic regimes corresponding to weak and strong effective
baryonic occupation.

Since the lattice-QCD equation of state is usually obtained
near vanishing baryon chemical potential, the most natural
and physically controlled expansion is the small-\(\mu/T\)
expansion. Therefore, in the present work the regime
\[
\frac{\mu(T)}{T}\ll1
\]
is treated as the main phenomenological approximation.
The opposite limit,
\[
\eta(T)=\frac{T}{T_H^2}\mu(T)<<1,
\]
may be considered as a complementary exploratory
parameterization relevant to strongly baryon-rich matter,
but it is less constrained by the present data set.

Thus, the function \(g(T)\) determines the basic temperature
evolution of nonconformality (the transition from the hadronic to the
quark-gluon regime), whereas the parameter \(B\) controls
the strength of the corrections associated with a finite chemical potential.

This expression reveals the key mechanism:
\begin{itemize}
\item $g(T)$ is an effective nonconformal spectral weight,
\item $\mu(T)$ controls their effective occupation,
\item the peak arises as a consequence of the interplay of these two factors.
\end{itemize}

This expression should be understood as an effective parameterization. It is intended
to reproduce lattice-QCD thermodynamics, and not as a rigorous
derivation from first-principles QCD.

\subsection{Comparison with lattice QCD}

The parameters are determined from a $\chi^2$ fit. The obtained curve reproduces:
\begin{itemize}
\item the position of the maximum ($T\sim200$ MeV),
\item the asymmetric shape,
\item the high-temperature tail.
\end{itemize}

\begin{figure}[t]
\centering
\includegraphics[width=0.85\linewidth]{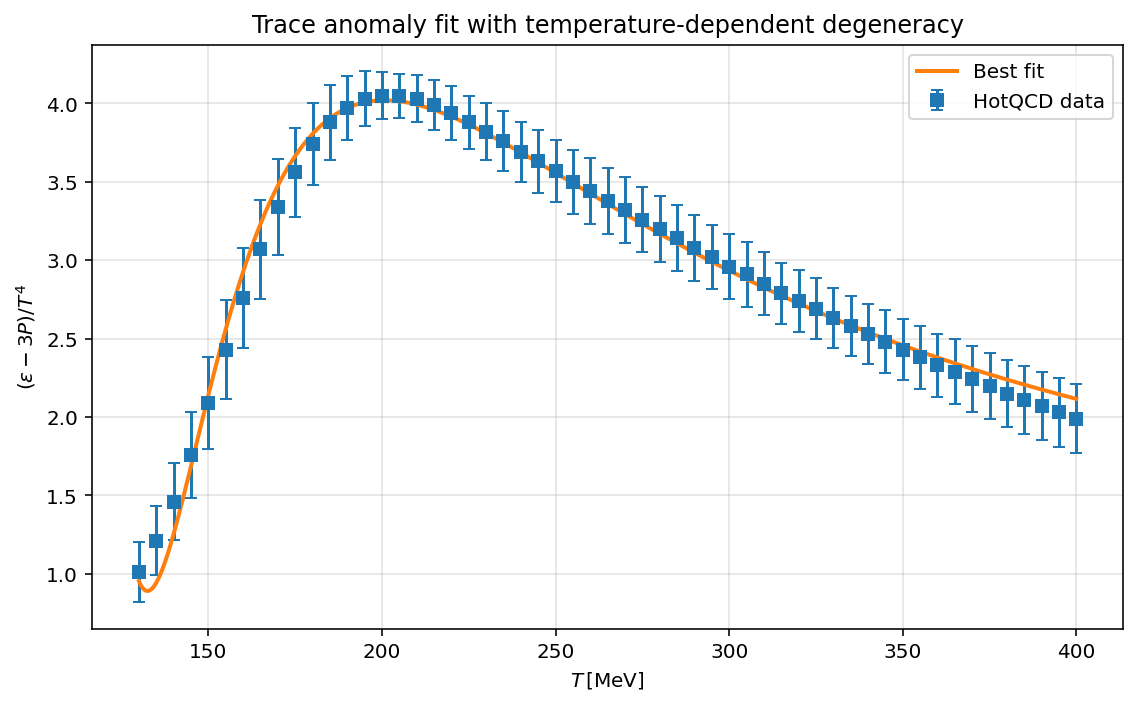}
\caption{
Trace anomaly in the regime
\(
\frac{\mu(T)}{T}\ll1
\).
The solid curve corresponds to the phenomenological model
with temperature-dependent effective degeneracy \(g(T)\),
while the points represent HotQCD lattice data.
}
\end{figure}

\begin{figure}[t]
\centering
\includegraphics[width=0.85\linewidth]{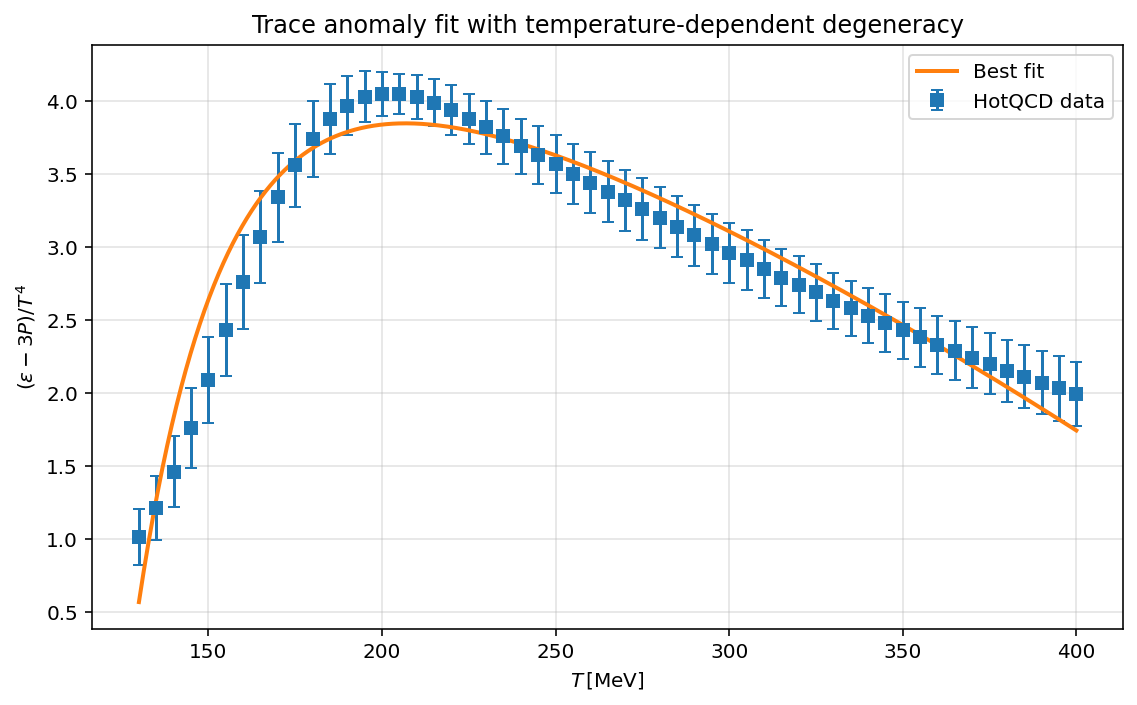}
\caption{
Trace anomaly obtained using the alternative
mixed thermal--chemical scaling parameterization.
The solid curve corresponds to the phenomenological model
with temperature-dependent effective degeneracy \(g(T)\),
while the points represent HotQCD lattice data.
}
\end{figure}

As a result, one obtains
\[
\chi^2 = 5.56,\qquad \chi^2_{\rm red}=0.11.
\]
The small value of \(\chi^2_{\rm red}\) indicates that the chosen
parameterization reproduces the main shape of the lattice data;
at the same time, this value should be interpreted with caution, since it
may also reflect conservative estimates of the errors in the input data set.

At the same time, in the regime
\[
\eta(T)=\frac{T}{T_H^2}\mu(T) <<1,
\]
the agreement becomes noticeably worse:
\[
\chi^2 = 40.22,
\qquad
\chi^2_{\mathrm{red}}\approx0.83.
\]

In the present fit the effective degeneracy varies
only moderately with temperature.
This may indicate that the dominant contribution
to the trace anomaly near the crossover is may indicate that, within the present phenomenological parameterization, 
the temperature dependence associated with the effective chemical-potential sector plays an important role in reproducing the lattice trace anomaly.

The fit drives the parameter $\alpha$ toward
very small values, indicating that within the
present phenomenological realization the lattice
trace anomaly can already be reproduced without
a strong explicit Hagedorn enhancement of $g(T)$.

The fit drives the parameter \(\alpha\) toward very small
values, indicating that within the present phenomenological
realization the lattice trace anomaly can already be reproduced
without a strong explicit Hagedorn enhancement of \(g(T)\).

An important result of the fit is that the lattice trace anomaly
can be reproduced even when the explicit temperature dependence
of the effective degeneracy \(g(T)\) becomes very weak.
This indicates that a substantial part of the nontrivial thermal
behavior is already generated by the dimensionless ratio
\[
\mu_{\rm eff}(T)/T,
\]
whose intrinsic \(1/T\)-type structure produces a strongly
nonlinear temperature dependence of the interaction measure.

\subsection{Physical interpretation}

The QCD crossover is determined by the competition of two effects:
\begin{enumerate}
\item the rapid growth of the number of degrees of freedom $g(T)$,
\item the suppression of the effective chemical potential $\mu(T)$.
\end{enumerate}

The peak of the trace anomaly arises precisely from this competition, giving a simple
and physically transparent interpretation, consistent with lattice QCD.

It arises as a consequence of the competition
of two opposite effects:

\[
\frac{d g(T)}{dT} > 0, \quad \frac{d \mu(T)}{dT} < 0.
\]

In the high-temperature limit, an approach
to the perturbative-QCD regime is expected, where
\[
\Delta(T) \to 0,
\]
which is consistent with asymptotic freedom.

\subsection{Hagedorn spectral density and effective degeneracy}

A complementary derivation of $g(T)$ follows from the Hagedorn picture
of the hadronic spectrum. Near the Hagedorn temperature the density of hadronic states approximately grows as
\begin{equation}
\rho(m)\simeq C\,m^{-a}\exp\left(\frac{m}{T_H}\right),
\end{equation}
where $C$ and $a$ are phenomenological constants~\cite{Hagedorn1965, Cleymans2006}.

The hadronic contribution to the thermodynamic weight then has the form
\begin{equation}
g_{\mathrm{had}}(T)
\sim
\int_0^\infty dm\,\rho(m)\exp\left(-\frac{m}{T}\right).
\end{equation}
Substituting the Hagedorn density, we obtain
\begin{equation}
g_{\mathrm{had}}(T)
\sim
\int_0^\infty dm\,m^{-a}
\exp\left[-m\left(\frac{1}{T}-\frac{1}{T_H}\right)\right].
\end{equation}

For $T<T_H$ this integral is finite, but it grows rapidly as
$T\rightarrow T_H^-$. Using
\begin{equation}
\int_0^\infty dm\,m^\nu e^{-\kappa m}
=\frac{\Gamma(\nu+1)}{\kappa^{\nu+1}},
\qquad \kappa>0,
\end{equation}
this suggests the qualitative behavior
\begin{equation}
g_{\mathrm{had}}(T)
\propto
\left(\frac{1}{T}-\frac{1}{T_H}\right)^{a-1}.
\end{equation}

In the present work the singular behavior is replaced
by a smooth phenomenological interpolation in order
to mimic finite-width effects, interactions, and the
crossover nature of the transition.

In compact phenomenological form this rapid growth can be represented as
\begin{equation}
g_{\mathrm{had}}(T)=
g_0\left[1+\alpha\left(\frac{T}{T_H}\right)^2\right],
\qquad T<T_H.
\end{equation}
Above the crossover region, deconfined partonic
degrees of freedom are gradually activated:
\begin{equation}
g_{\mathrm{qgp}}(T)=
g_\infty\left[1-e^{-\beta(T-T_H)}\right],
\qquad T\ge T_H.
\end{equation}

Thus, the adopted parameterization of $g(T)$ can be understood as a smooth effective
representation of two physical mechanisms: the Hagedorn growth
of hadronic resonances below $T_H$ and the activation of quark-gluon degrees
of freedom above the crossover.

From the formula
\[
g_{\rm had}(T)\sim
\int_0^\infty dm\, m^{-a}
\exp\left[-m\left(\frac{1}{T}-\frac{1}{T_H}\right)\right]
\]
it is seen that, when approaching the Hagedorn temperature,
\[
\kappa(T)=\frac{1}{T}-\frac{1}{T_H}\to 0^+ .
\]

Therefore the Hagedorn contribution grows rapidly:
\begin{equation}
g_{\rm had}(T)\sim \mathcal{F}\!\left(\kappa(T)\right),
\qquad
\kappa(T)=\frac{1}{T}-\frac{1}{T_H}\to 0^+,
\end{equation}

where the specific character of the growth depends on the power-law
prefactor in the spectral density and on the statistical weight of
individual resonances. In any case the exponential factor
\(\exp(m/T_H)\) leads to a sharp increase of the spectral weight
near \(T_H\).

In an idealized model with an infinite spectrum this can lead
to a singular growth of the effective degeneracy as
\(T\to T_H^-\). In a real system this behavior is smoothed out by
interactions, the finite widths of resonances, and the onset of
deconfinement.

It is tempting to speculate that the effective temperature dependence of $g(T)$ may be related to the renormalization-group evolution of thermodynamically active QCD channels. In this sense, 
One may speculate that renormalization-group evolution and operator anomalous dimensions effectively contribute to the thermal redistribution of spectral weight $g(T)$ \cite{TiulupaKrivenkoEmetov2026}. 
However, such an interpretation remains heuristic and is beyond the scope of the present work.

\section{Conclusions}

In this work a compact analytical scheme has been proposed for describing
strongly interacting matter in the QCD-crossover region. The approach is based
on a multicomponent van der Waals model and is supplemented by
a temperature-dependent effective degeneracy $g(T)$ and an effective
chemical potential $\mu(T)$, which makes it possible to account for the change of
thermodynamically active degrees of freedom and their occupation.

The key result is that the trace anomaly
\begin{equation}
\Delta(T)=\frac{\varepsilon-3P}{T^4}
\end{equation}
can be represented in the form
\begin{equation}
\Delta(T)=g(T)\left[1+B\left(\frac{\mu(T)}{T}\right)^2\right],
\end{equation}
where $g(T)$ effectively describes the number of active degrees of freedom,
and $\mu(T)$ describes the degree of their thermal occupation and the processes
of melting of bound states.

Two phenomenological parameterizations of the
trace anomaly were investigated.

The first one is based on the standard
small-\(\mu/T\) thermodynamic expansion and
provides an accurate description of the lattice data.

The second one employs a mixed thermal-chemical
scaling variable
\[
\eta(T)=\frac{T}{T_H^2}\mu(T),
\]
and should be regarded as an alternative effective
phenomenological ansatz.

This demonstrates that the proposed phenomenological
scheme naturally interpolates between nearly baryon-symmetric
matter and strongly baryon-rich QCD matter.

Comparison with HotQCD lattice data demonstrates that
the proposed parameterization provides a particularly accurate
description of the weakly baryonic near-crossover regime.
For this region the fit yields
\[
\chi^2 = 5.56,
\qquad
\chi^2_{\mathrm{red}} \approx 0.11.
\]

The model correctly reproduces both the position of the maximum of
\(\Delta(T)\) at \(T \approx 200\) MeV and the characteristic asymmetric
shape of the trace anomaly curve.

A substantial difference between the two asymptotic
regimes has been observed.

In the region
\[
\frac{\mu(T)}{T}\ll1,
\]
the model provides an accurate description of the lattice
trace anomaly, yielding
\[
\chi^2 = 5.56,
\qquad
\chi^2_{\mathrm{red}}\approx0.11.
\]

At the same time, in the regime

\[
\eta(T)=\frac{T}{T_H^2}\mu(T)  \ll 1,
\]

the agreement becomes noticeably worse:
\[
\chi^2 = 40.23,
\qquad
\chi^2_{\mathrm{red}}\approx0.83.
\]

Numerically, this parameterization provides a
noticeably worse description of the lattice data
than the standard small-\(\mu/T\) expansion.

This may indicate that the dominant nonconformal
corrections near the lattice-QCD crossover are
more naturally organized in powers of the ratio
\(\mu/T\), whereas mixed thermal-chemical scaling
variables of the form \(T\mu/T_H^2\) capture only
part of the underlying dynamics.

The substantially better agreement obtained within
the standard small-\(\mu/T\) expansion may indicate
that the dominant nonconformal corrections near
the QCD crossover are primarily controlled by the
ratio \(\mu/T\).

This indicates that the strongly baryon-rich regime
likely contains additional nonperturbative effects
which are not fully captured within the simplest
effective parameterization employed in the present work.

This suggests that the dense baryonic regime possesses
a more complicated nonperturbative structure than the
near-crossover weakly baryonic sector.

This behavior may indicate that, in the region
of strong effective baryonic occupation,
the thermodynamics becomes increasingly sensitive
to correlation effects, residual confinement dynamics,
and possible medium modifications of the effective
degrees of freedom.

Therefore, the dense baryonic regime may require
a more complicated dynamical structure of the
effective spectral weight \(g(T)\) and of the
chemical-potential sector.

An important result of the fit is that the parameter
\(\alpha\), controlling the explicit temperature dependence
of the effective degeneracy \(g(T)\), is systematically
driven toward values numerically close to zero.
This may indicate that the dominant contribution to the
trace anomaly near the crossover is governed primarily
by the effective chemical-potential sector rather than by
a strong explicit Hagedorn-type growth of thermodynamic
degrees of freedom within the present ansatz.

At the same time, the introduction of the effective
chemical potential \(\mu(T)\) allows the model
to be generalized beyond the nearly baryon-symmetric
lattice-QCD regime toward dense baryonic matter,
which may be relevant for heavy-ion collisions
at intermediate energies and for compact-star physics.

The small value of $\chi^2_{\mathrm{red}}$ indicates that the main
interaction effects can be effectively accumulated in the function $g(T)$,
which combines the contribution of the Hagedorn growth of the density of states
and the activation of quark-gluon degrees of freedom.

This indicates that the dominant nonconformal effects
of strongly interacting matter can be effectively encoded
in the temperature evolution of the spectral weight \(g(T)\)
and in the occupation dynamics governed by \(\mu(T)\).

The proposed approach should be regarded as an effective
phenomenological description that generalizes models of multicomponent
hadronic matter
\cite{Krivenko2017Attraction,Krivenko2019,Krivenko2023}
and is consistent with the results of lattice QCD
\cite{Bazavov2014,Borsanyi2014}, but is not a rigorous derivation
from first-principles QCD.

The functional form of $g(T)$ has a natural physical interpretation.
On the one hand, within the statistical Hagedorn model,
the exponential growth of the spectral density leads to a
rapid increase in the number of hadronic states as $T \to T_H$.
On the other hand, in QCD the effective nonconformal spectral weight
may be associated with the spectrum of anomalous dimensions and
renormalization-group evolution of operators, which determines their contribution
to thermodynamics.

Within this picture, the QCD crossover is interpreted as the result
of the competition of two mechanisms:
\begin{enumerate}
\item the growth of the number of active hadronic and partonic degrees of freedom;
\item the suppression of the effective chemical potential,
associated with deconfinement and the destruction of bound states.
\end{enumerate}
Their interplay naturally leads to the formation of the maximum of
the trace anomaly.

Thus, the proposed model establishes a connection between
the van der Waals dynamics of hadronic matter,
the Hagedorn spectrum, the renormalization-group motivated evolution of QCD,
and the results of lattice calculations, providing a compact
analytical description of the crossover.

Further development of the approach may include a microscopic
derivation of $g(T)$ from the spectral density, inclusion of
the dynamics of the Polyakov loop, temperature-dependent quasiparticle
masses, as well as a generalization to transport coefficients and
the speed of sound.

Such an interpretation suggests that the QCD crossover
may be understood not merely as a transition between
hadronic and quark-gluon phases, but as a continuous
redistribution of nonconformal spectral weight between
different thermodynamically active channels.

Within this picture, the peak of the trace anomaly emerges
as a dynamical consequence of the competition between
the activation of new degrees of freedom and the suppression
of effective baryonic occupation.

The observed asymmetry between the regimes
\(
\mu(T)/T \ll 1
\)
and $\eta(T)=\frac{T}{T_H^2}\mu(T)   \ll 1,$
may itself represent an important physical signal of the
changing microscopic structure of strongly interacting matter
across the QCD phase diagram.

The present model should be regarded as a compact
effective phenomenological parameterization.
In particular, the functions \(g(T)\) and \(\mu(T)\)
are not derived microscopically from QCD,
but effectively encode the dominant thermodynamic
nonconformal contributions near the crossover region.

This may provide a useful phenomenological framework
for connecting lattice-QCD thermodynamics,
dense hadronic matter, and effective descriptions
of strongly interacting systems at finite temperature
and baryon density.

The present model does not include explicit confinement dynamics,
Polyakov-loop effects, or a microscopic derivation of the effective
functions \(g(T)\) and \(\mu(T)\). Therefore the proposed scheme
should be regarded as a compact phenomenological description of
nonconformal thermodynamics near the QCD crossover.

\section*{Acknowledgments}

The author is grateful to colleagues from the Institute for Nuclear Research of the National Academy of Sciences of Ukraine and the National Technical University of Ukraine “Igor Sikorsky Kyiv Polytechnic Institute” for valuable discussions concerning the thermodynamics of strongly interacting matter, QCD crossover physics, trace anomaly phenomenology, and effective descriptions of temperature-dependent degrees of freedom in quark--gluon and hadronic matter.


@book{Shuryak2004,
  author    = {Shuryak, E. V.},
  title     = {The QCD Vacuum, Hadrons and Superdense Matter},
  publisher = {World Scientific},
  address   = {Singapore},
  year      = {2004},
  edition   = {2},
  doi       = {10.1142/5367}
}

@article{Gyulassy2005,
  author  = {Gyulassy, M. and McLerran, L.},
  title   = {New forms of QCD matter discovered at RHIC},
  journal = {Nucl. Phys. A},
  volume  = {750},
  year    = {2005},
  pages   = {30--63},
  doi     = {10.1016/j.nuclphysa.2004.10.034},
  eprint  = {nucl-th/0405013},
  archivePrefix = {arXiv}
}

@article{BraunMunzinger2007,
  author  = {Braun-Munzinger, P. and Stachel, J.},
  title   = {The quest for the quark--gluon plasma},
  journal = {Nature},
  volume  = {448},
  year    = {2007},
  pages   = {302--309},
  doi     = {10.1038/nature06080}
}

@article{Krivenko2023,
  author        = {Krivenko-Emetov, Yaroslav D.},
  title         = {Multicomponent van der Waals Model of a Nuclear Fireball in the Freeze-Out Stage},
  journal       = {Letters in High Energy Physics},
  volume        = {2023},
  year          = {2023},
  pages         = {401},
  doi           = {10.31526/LHEP.2023.401},
  eprint        = {2301.00742},
  archivePrefix = {arXiv},
  primaryClass  = {hep-ph}
}

@article{Gorenstein1999,
  author        = {Gorenstein, M. I. and Kostyuk, A. P. and Krivenko, Ya. D.},
  title         = {Van der Waals excluded-volume model of multicomponent hadron gas},
  journal       = {J. Phys. G: Nucl. Part. Phys.},
  volume        = {25},
  number        = {9},
  year          = {1999},
  pages         = {L75--L83},
  doi           = {10.1088/0954-3899/25/9/102},
  eprint        = {nucl-th/9906068},
  archivePrefix = {arXiv}
}

@article{Rischke1991,
  author  = {Rischke, D. H. and Gorenstein, M. I. and Stoecker, H. and Greiner, W.},
  title   = {Excluded volume effect for the nuclear matter equation of state},
  journal = {Z. Phys. C},
  volume  = {51},
  year    = {1991},
  pages   = {485--490},
  doi     = {10.1007/BF01548574}
}

@inproceedings{TiulupaKrivenkoEmetov2026,
  author    = {O. Tiulupa and Y. D. Krivenko-Emetov},
  title     = {Multicomponent Model of Strongly Interacting Matter Near the QCD Crossover},
  booktitle = {Proc. XXIV All-Ukrainian Scientific and Practical Conference of Students, Postgraduates and Young Scientists ``Theoretical and Applied Problems of Physics, Mathematics and Informatics''},
  year      = {2026},
  address   = {Kyiv, Ukraine},
  pages     = {75--80},
  note      = {in Ukrainian}
}

@article{Vovchenko2017,
  author  = {Vovchenko, Volodymyr and Motornenko, Anton and Alba, Paolo and Gorenstein, Mark I. and Satarov, Leonid M. and Stoecker, Horst},
  title   = {Multicomponent van der Waals equation of state: Applications in nuclear and hadronic physics},
  journal = {Phys. Rev. C},
  volume  = {96},
  year    = {2017},
  pages   = {045202},
  doi     = {10.1103/PhysRevC.96.045202}
}

@inproceedings{Krivenko2017Attraction,
  author    = {Krivenko-Emetov, Ya. D.},
  title     = {Interparticle attractive forces account of the multicomponent hadron gas in the grand canonical ensemble},
  booktitle = {Proceedings of the XXIV Annual Scientific Conference of the Institute for Nuclear Research of the National Academy of Sciences of Ukraine},
  address   = {Kyiv, Ukraine},
  year      = {2017},
  pages     = {36},
  publisher = {INR NAS of Ukraine},
  note      = {in Ukrainian}
}

@article{Krivenko2019,
  author        = {Krivenko-Emetov, Ya. D.},
  title         = {Attractive inter-particle force in Van der Waals model of hadron gas in the grand canonical ensemble},
  journal       = {arXiv},
  year          = {2019},
  eprint        = {1909.08441},
  archivePrefix = {arXiv},
  primaryClass  = {hep-ph}
}

@article{Borsanyi2014,
  author        = {Borsanyi, S. and Fodor, Z. and Hoelbling, C. and Katz, S. D. and Krieg, S. and Szabo, K. K.},
  title         = {Full result for the QCD equation of state with 2+1 flavors},
  journal       = {Phys. Lett. B},
  volume        = {730},
  year          = {2014},
  pages         = {99--104},
  doi           = {10.1016/j.physletb.2014.01.007},
  eprint        = {1309.5258},
  archivePrefix = {arXiv},
  primaryClass  = {hep-lat}
}

@article{Bazavov2014,
  author        = {Bazavov, A. and Bhattacharya, T. and DeTar, C. and Ding, H.-T. and Gottlieb, Steven and Gupta, Rajan and Hegde, P. and Heller, U. M. and Karsch, F. and Laermann, E. and Levkova, L. and Mukherjee, Swagato and Petreczky, P. and Schmidt, C. and Schroeder, C. and Soltz, R. A. and Soeldner, W. and Sugar, R. and Wagner, M. and Vranas, P.},
  title         = {Equation of state in (2+1)-flavor QCD},
  journal       = {Phys. Rev. D},
  volume        = {90},
  year          = {2014},
  pages         = {094503},
  doi           = {10.1103/PhysRevD.90.094503},
  eprint        = {1407.6387},
  archivePrefix = {arXiv},
  primaryClass  = {hep-lat}
}

@article{Hagedorn1965,
  author  = {Hagedorn, R.},
  title   = {Statistical thermodynamics of strong interactions at high energies},
  journal = {Nuovo Cim. Suppl.},
  volume  = {3},
  year    = {1965},
  pages   = {147--186}
}

@book{Kapusta2006,
  author    = {Kapusta, J. I. and Gale, C.},
  title     = {Finite-Temperature Field Theory: Principles and Applications},
  publisher = {Cambridge University Press},
  address   = {Cambridge},
  year      = {2006},
  edition   = {2nd},
  isbn      = {9780521820820},
  doi       = {10.1017/CBO9780511535130}
}

@article{Cleymans2006,
  author        = {Cleymans, J. and Oeschler, H. and Redlich, K. and Wheaton, S.},
  title         = {Comparison of chemical freeze-out criteria in heavy-ion collisions},
  journal       = {Phys. Rev. C},
  volume        = {73},
  year          = {2006},
  pages         = {034905},
  doi           = {10.1103/PhysRevC.73.034905},
  eprint        = {hep-ph/0511094},
  archivePrefix = {arXiv}
}
\end{document}